\documentclass[twocolumn,letter]{jpsj2} 
\newcommand{\rC}{{\rm CP}}

\newcommand{\rS}{{\rm SB}}
\usepackage{bm}

\def\runauthor{Yuki \textsc{Fuseya}
{\it et al}.
}
\title{
Effect of Inter-Site Repulsions 
on  Magnetic Susceptibility of 
One-Dimensional Electron Systems at Quarter-Filling 
}

\author{Yuki \textsc{Fuseya}
\thanks{E-mail address: fuseya@slab.phys.nagoya-u.ac.jp.}, 
Masahisa \textsc{Tsuchiizu}, 
Yoshikazu \textsc{Suzumura} 
and Claude \textsc{Bourbonnais}$^{1}$}

\inst{Department of Physics, Nagoya University, Nagoya 464-8602 \\
$^{1}$Regroupement Qu\'{e}becois sur les Mat\'{e}riaux de Pointe, D\'{e}partement de physique, Universit\'{e} de Sherbrooke, Sherbrooke, Qu\'{e}bec, Canada, J1K-2R1
}

\abst{
   The temperature dependence of the magnetic susceptibility, $\chi (T)$, is investigated for one-dimensional interacting electron systems at quarter-filling within the Kadanoff-Wilson renormalization-group method. 
   The forward scattering on the same branch (the $g_4$-process) is  examined together with the backward ($g_1$) and forward ($g_2$) scattering amplitudes on opposite branches.
   In connection with lattice models, we show that $\chi (T)$ is strongly enhanced by the nearest-neighbor interaction, an enhancement that surpasses one of the next-nearest-neighbor interaction.
   A connection between our predictions for  $\chi (T)$ and experimental results for $\chi (T)$ in quasi-one-dimensional organic conductors is presented.
}

\kword{
magnetic susceptibility, renormalization group, long-range Coulomb interaction, extended-Hubbard model, quarter-filling, one-dimension, (TMTTF)$_2 $X, (TMTSF)$_2 $X
}

\begin{document}
\maketitle

   In low dimensional conductors, the influence of long-range Coulomb interactions, or the off-site interaction $V$'s are known to lead to various interesting ordered states, such as charge ordering\cite{SHF} and spin-density wave (SDW) state coexisting with charge-density wave (CDW)\cite{PR,KOY}. 
   At the present stage, however, the effect of $V$'s is not clearly understood compared with that of the on-site repulsion $U$. 
   The purpose of this letter is to analyze the effect of $V$'s on the quasi-one-dimensional conductor in the normal state by calculating the temperature dependence of the magnetic susceptibility $\chi (T)$ for the one-dimensional system at quarter-filling. 
   The results shed some light on the magnetic properties of both normal and ordered states in these kinds of materials.
   
   A lot of studies of $\chi (T)$ were devoted to the one-dimensional (1D) Hubbard model, i.e., for $V$'s $=0$.
   The Bethe ansatz gives us an exact solution, but $\chi$ only at $T=0$ is available for the 1D Hubbard model\cite{Shiba}.
   The temperature dependence of $\chi$ can be extracted from numerical simulations but only at high temperatures\cite{NBTVT}.
   For $U$ smaller than the bandwidth, the renormalization-group (RG) approach gives results that quantitatively agree with both the exact solution at $T=0$ and the numerical solutions at high temperatures\cite{NBTVT}.
   In the presence of $V$'s, however, there is no exact solution, and  the size of numerical simulations becomes exceedingly large.
   These techniques are then of limited use to investigate the effect of $V$'s.

   In this letter, we use the RG technique to calculate $\chi (T)$ for the full temperature range from zero to the bandwidth $E_0$ in the presence of $V$'s ($>0$).
   Considering the possible interactions for branches of right and left going electrons in the continuum limit, we take into account the forward scattering on the same branch (SB), denoted as the $g_4$-process, with the backward ($g_1$) and forward ($g_2$) scattering processes on opposite branches. 
   This is the first derivation of the RG flows of the sets of coupling constants and $\chi (T)$ at the one-loop level including the non-logarithmic channels for particles on the SB, which become important at finite-temperatures.

   Our main results are:  
   1) $\chi (T)$ is noticeably enhanced at finite temperature by the nearest-neighbor repulsion ($V_1$) compared with the next-nearest-neighbor repulsion ($V_2$);
   2) $\chi (T=0)$ is not simply proportional to the inverse of the ordinary spin-velocity $v_\sigma$;
   3)  $\chi (T)$ is reduced by a moderately large $V_2$ due to the CDW fluctuations.

   We consider the quarter-filled extended-Hubbard model 
   in 1D with the Hamiltonian  $H=H_0 + H_{\rm I}$, where 
\begin{align}
   H_0 &= -t\sum_{i, \sigma}
   \left[ c_{i+1, \sigma}^\dagger c_{i, \sigma}^{} + {\rm h. c.}
 \right],
\label{H0}
\\
   H_{\rm I}
   &=\sum_{i}\left[
   U n_{i \uparrow}  n_{i \downarrow}
   +V_1 n_{i}  n_{i+1}
   +V_2 n_{i}  n_{i+2}
   \right].
   \label{HIHub}
\end{align}
   Here $t$ denotes the intrachain hopping integral, $c_{i, \sigma}^{(\dagger)}$ as an annihilation (a creation) operator on the $i$-th site with spin $\sigma$ $(=\uparrow, \downarrow)$. 
   The density operators are $n_{i, \sigma}=c_{i, \sigma}^\dagger c_{i, \sigma}$ and $n_i=n_{i \uparrow}+n_{i \downarrow}$.
   In weak-coupling regions, Eq. (\ref{H0}) can be safely expressed as
\begin{align}
   H_0 &= \sum_{p, k, \sigma} \epsilon_p (k) 
   c_{p, k, \sigma}^\dagger c_{p, k, \sigma}, 
\end{align}
   where $\epsilon_p (k) =v_F (pk-k_F )$ is the linearized 1D dispersion and $v_F$ being the Fermi velocity.
   %
   %
   The operator $c_{p, k, \sigma}^{(\dagger)}$ annihilates (creates) an electron of spin $\sigma$ close to the  Fermi point of the right $k=+k_F$ ($p=+1$) and left $k=-k_F$ ($p=-1$) branches.
   We focus on the scattering processes between electrons near the Fermi points and express the Hamiltonian in terms of the charge- and spin-couplings in the following form:
   %
\begin{align}
   H_{\rm I}
   =&\frac{1}{L} \sum_{q, p}
   \bigl[
   g_{\rho}
   \rho_p (q) \rho_{-p} (-q)
   +g_{\sigma}
   \bm{S}_p (q)\cdot \bm{S}_{-p} (-q)
   \nonumber\\&
   +g_{4 \rho}
   \rho_p (q) \rho_p (-q)
   +g_{4\sigma}
   S^z_p (q)\cdot S^z_{p} (-q)
   \bigr] ,
   \label{HIrs}
\end{align}
where
\begin{align}
   \rho_p (q)
   &\equiv \frac{1}{2}\sum_{k, \sigma}
   c_{p,k+q,\sigma}^\dagger
   c_{p,k,\sigma}^{}, \\
   \bm{S}_p (q)
   &\equiv \frac{1}{2}\sum_{k, \alpha, \beta}
   c_{p,k+q,\alpha}^\dagger
   \pmb{\sigma}^{\alpha \beta}
   c_{p,k\beta}^{},
\end{align}
   are the charge and spin-density of the branch $p$, respectively.
   Here $\pmb{\sigma}$ is the vector of the Pauli matrices.
   The $g_{4\sigma}$-term is equivalent to $(g_{4\sigma}/3)\bm{S}_p (q)\cdot \bm{S}_p (q)$ due to the spin-rotational SU(2) symmetry of the original extended-Hubbard Hamiltonian.
   The coupling constants at 1/4 filling ($k_F =\pi /4, v_F =\sqrt{2}t$) are given by
\begin{align}
   g_{\rho}&=U + 4V_1 + 6V_2 ,\\
   g_{\sigma}&=-U + 2V_2 , \\
   g_{4\rho}&= U + 4V_1 + 4V_2 , \\
   g_{4\sigma}&= -U.
   \label{g4sUV}
\end{align}
   The lattice constant is taken as unity.
   %

	\begin{figure}[tb]
	\begin{center}\leavevmode
   \includegraphics[width=8cm]{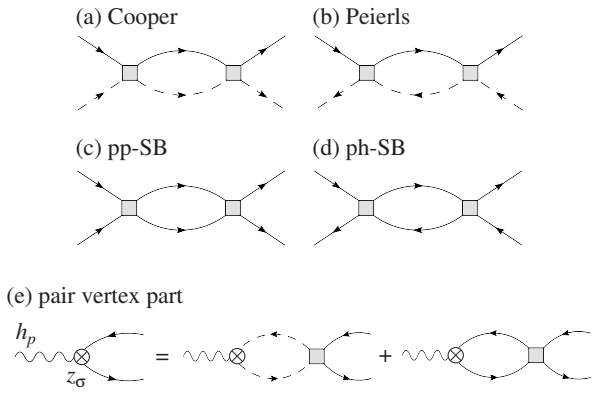}
	\end{center}
	\caption{
   Diagrams at one-loop level for (a) Cooper and (b) Peierls channels, and (c) pp and (d) ph channels on the same branch (SB); (e) perturbative expansion of the pair vertex part $z_\sigma$ [see eq. (\ref{dz})].
   The solid (dashed) line indicates electrons for right (left) going state, and the shaded square denotes respective coupling constants. 
   }
	\label{CPL}
	\end{figure}

   Following the Kadanoff-Wilson RG technique\cite{BC}, each RG step 
   consists of the partial integrations of the fermion degrees of freedom 
   in the outer band-momentum shell,  $E_0 (\ell) /2 \geqq \epsilon_p (k) > E_0 (\ell +d\ell)/2 $ for electrons and $-E_0 (\ell) /2 \leqq \epsilon_p (k) < -E_0 (\ell +d\ell)/2 $ for holes.
   Here, $E_0 (\ell )$ is the renormalized bandwidth $E_0 (\ell )=E_0 e^{-\ell}$ with the initial bandwidth $E_0 =2v_F k_F \simeq 2t$.
   At the one-loop level (Fig.\ref{CPL}), RG equations for coupling constants at finite temperature are given by 
   \begin{align}
   \frac{d}{d\ell}G_\rho(\ell) &=
   -\frac{1}{4}G_\rho(\ell)
   \left[G_{4\rho}(\ell)-G_{4\sigma}(\ell)\right]
   I_\rS (\ell), \label{dl1}
   \\
   \frac{d}{d\ell}G_\sigma(\ell)
   &=G_\sigma^2 (\ell)I_\rC (\ell)
   \nonumber\\&
   +\frac{1}{4}G_\sigma(\ell) 
   \left[ G_{4\rho}(\ell )-G_{4\sigma}(\ell)\right]
   I_\rS(\ell),\label{dl2}
   \\
   \frac{d}{d\ell}G_{4\rho}(\ell)&=-\frac{1}{2}
   \left[G_\rho^2(\ell) -
   G_{4\rho}(\ell)G_{4\sigma}(\ell)\right]I_\rS(\ell),\label{dl3}
   \\
   \frac{d}{d\ell}G_{4\sigma}(\ell)&=-\frac{1}{2}
   \left[3G_\sigma^2(\ell) -
   G_{4\rho}(\ell)G_{4\sigma}(\ell)\right]I_\rS(\ell),
\label{dl4}
\end{align}
   where $G_{\nu}(\ell)$'s are dimensionless couplings 
   with the initial conditions  $G_{\nu}(0)= g_{\nu}/\pi v_F$.
   The quantities
   $I_\rC (\ell)= \tanh [E_0 (\ell)/4T]$
   are the Cooper/Peierls outer shell contractions [Fig. \ref{CPL} (a), (b)], which are cut off by the temperature, namely 
   $ I_\rC (\ell) \simeq 1 $ for $E_0 (\ell)/4 \gtrsim T$ and $ I_\rC (\ell) \simeq  0$ otherwise.

   The function $I_\rS (\ell)$ is the coefficient of the particle-particle (pp) and particle-hole (ph) contractions on the SB [Fig. \ref{CPL} (c) and (d), respectively]
   given by 
\begin{align}
   I_\rS (\ell) d\ell &= -\lim_{q\to 0}\frac{2\pi v_F T}{L}
   \sum_{\omega_n}\sum_{k}^{\rm shell}
   \mathfrak{G}_p^0(k, \omega_n )\mathfrak{G}_p^0 (k+q, \omega_n)
   \nonumber\\&
   =\frac{E_0 (\ell )}{4T}
   \cosh^{-2}\frac{E_0 (\ell )}{4T} d\ell.
   \label{IL}
\end{align}
   where $\mathfrak{G}_p^0 (k, \omega_n )=\left[ {\rm i}\omega_n - \epsilon_p (k) \right] ^{-1}$ is the non-interacting Green's function
    with a Matsubara frequency $\omega_n$ . 
   The SB channels become finite due to thermal particle-hole excitations, i.e.,  $E_0 (\ell)/4 \sim T$.

   By setting $I_\rS (\ell )=0$ and $I_\rC (\ell)=1$, 
   Eqs. (\ref{dl1})-(\ref{dl4}) reproduce the previous RG equations which treat only the Cooper and Peierls contractions.\cite{Solyom,BC} 
   Equations (\ref{dl1})-(\ref{dl4}) are the full RG equations at the one-loop level.
   Here, ``full" means they include not only the logarithmic (Cooper and Peierls) channels, but also the non-logarithmic (SB) channel.
   By considering the contribution of $I_{\rm SB}$, the renormalization of $g_{4\rho, 4\sigma}$ and their influence on the charge- and magnetic-susceptibility is obtained.
   %
   %
   %
   In Eqs. (\ref{dl1})-(\ref{dl4}), they apparently imply that the spin- and charge-part are coupled together.
   However, the contribution of the SB channel, $I_\rS (\ell )$, is small compared to that of the logarithmic term, $I_\rC (\ell)$, so that the coupling of spin- and charge-part is very weak.
   Further, at the ground state, the present flow equations exhibit the spin-charge decoupling since $I_\rS (\ell)=0$ at $T=0$.

   The uniform magnetic susceptibility $\chi(T)$ can be calculated by introducing a Zeeman coupling between a source field  $h_p(q)$ and the spin-density variable $\bm{S}_p(q)$.
   By adding $H_h = \sum_{q,p} z_\sigma h_p(q) \bm{S}_p(q)$ to the Hamiltonian, the uniform magnetic susceptibility can be obtained and which leads,
   \begin{align}
   \chi (T) = \frac{2}{\pi v_F}\int_{0}^\infty 
   \left[
   z_\sigma (\ell )
   \right] ^2 I_\rS (\ell)
   d\ell ,
   \label{magsusRG}
\end{align}
   where $z_{\sigma}(\ell)$ is the pair vertex part and is calculated from 
\begin{align}
   \frac{d}{d\ell}z_{\sigma}(\ell)
   =\frac{1}{2}z_{\sigma}(\ell ) 
   \left[
   -G_\sigma (\ell) +G_{4\perp}(\ell)
   \right] I_\rS (\ell),
   \label{dz}
\end{align}
  with 
\begin{align}
   G_{4\perp}(\ell) \equiv \frac{1}{2}
   [G_{4\rho}(\ell)-G_{4\sigma}(\ell)],
\end{align}
   and the initial condition $z_{\sigma}(0)=1$ [see Fig. \ref{CPL} (e)].

	\begin{figure}[tb]
	\begin{center}\leavevmode
   \includegraphics[width=8cm]{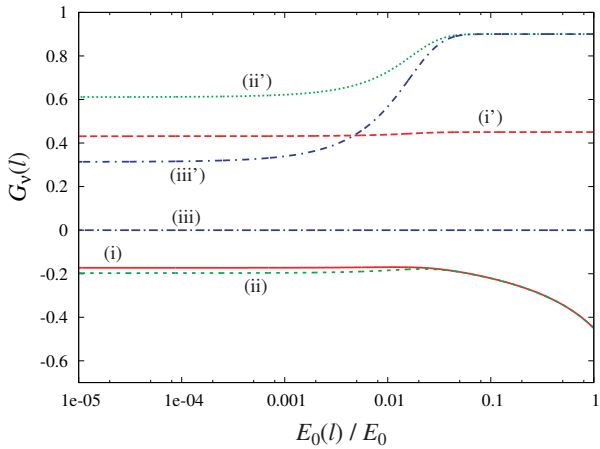}
	\end{center}
	\caption{Energy flow of the couplings at $T=0.01t$. (The SB channel becomes effective around $E_0 (\ell ) \sim 4T =0.018E_0$.)
   $G_\sigma$ and $G_{4\perp} (\equiv (G_{4\rho}-G_{4\sigma})/2)$ are shown by lines (i) and (i') for $(U, V_1, V_2 )=(2t, 0, 0)$, by lines (ii) and (ii') for $(U, V_1 , V_2 )=(2t, t, 0)$ and by lines (iii) and (iii') for $(U, V_1 , V_2)=(2t, 0, t)$.
   }
	\label{gflow}
	\end{figure}

   The  flows of the couplings as a function of energy  at $T=0.01t$ are shown in Fig. \ref{gflow}.
   We show results for the following different sets of parameters; 
   (i) $(U, V_1, V_2 )=(2t, 0, 0)$,
   (ii) $(U, V_1 , V_2 )=(2t, t, 0)$ and 
   (iii) $(U, V_1 , V_2)=(2t, 0, t)$.
   Note that for the cases (i) and (ii), the SDW becomes dominant, while both the SDW and CDW are enhanced for the case (iii).
   Generally, for $E_0 (\ell )/4 \gtrsim T$, $G_\sigma(\ell)$ follows $G_\sigma(\ell) = G_\sigma(0) / \left[ 1+G_\sigma(0) \ln (E_0 (\ell)/E_0 )\right]$, as found in the conventional RG.
   With decreasing $E_0 (\ell )$, $|G_\sigma |$'s [plots (i) \& (ii)] are gradually suppressed due to the irrelevance of $G_\sigma (<0)$ in the limit of $T=0$. 
   However, the $G_{4\perp}$'s [plots (i'), (ii') \& (iii')] keep their initial values until $E_0 (\ell )$ decreases to $E_0 (\ell )/4 \sim T$. 
   Around $E_0 (\ell )/4 \sim T$, the SB channel becomes effective while the contributions of the Cooper and Peierls channels to the flow are strongly suppressed.
   As a result, the SB channel slightly enhances $|G_\sigma |$ and remarkably reduces $G_{4\perp}$ [plots (ii) \& (ii')] for finite $V_1$.
   For finite $V_2$, $G_{4\perp}$ [plot (iii')] is reduced further.
   The contribution of the SB channel is relatively small for the Hubbard model ($V_{1,2}=0$) [plots (i) \& (i')].
   As shown later, the properties of $\chi (T)$ are directly affected  by the temperature dependence of $G_\sigma $ and $G_{4\perp}$.  
   %

	\begin{figure}[bt]
	\begin{center}\leavevmode
   \includegraphics[width=8cm]{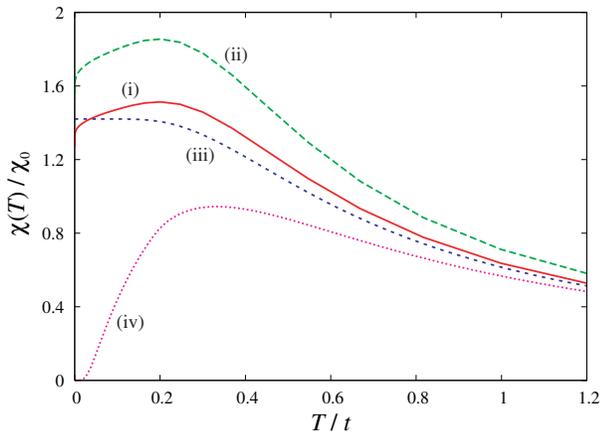}
	\end{center}
	\caption{
   Magnetic susceptibility $\chi (T)$ for (i) $(U, V_1, V_2)$=$(2t, 0, 0)$, (ii) $(2t, t, 0)$, (iii) $(2t, 0, t)$ and (iv) $(2t, 0, 2t)$, where $\chi_0 =2/\pi v_F (\simeq 0.45/t)$ denotes the magnetic susceptibility without interactions at $T=0$.  
   }
	\label{magsus}
	\end{figure}
   %
   
   The temperature dependence of the magnetic susceptibility $\chi (T)$ is shown in Fig. \ref{magsus}.
   The result for the Hubbard model ($U=2t$) given by the plot (i) is quantitatively consistent with $\chi (T)$ calculated by the quantum Monte Carlo (QMC) simulations\cite{NBTVT}.
   The value at $T=0$, $\chi (0)t=0.58$, agrees well with the exact solution, $\chi_{\rm exact} (0)t= 0.57$, for $U=2t$.\cite{Shiba} 
   Near zero temperature, $\chi (T)$ decreases as $d\chi /d T \to \infty$ for plots (i) and (ii), which has also been pointed out in the context of the Heisenberg spin chain\cite{EAT}.
   This logarithmic decrease of $\chi (T\sim 0)$ is due to the logarithmic decrease of $|G_\sigma|$.\cite{NBTVT}
   Note that for large $U$, e.g., $U\gtrsim 4t$, there was a sizable difference between the previous $\chi (T)$ of the RG approach \cite{NBTVT} and the $\chi(T)$ of the QMC calculation at high temperatures.
   The difference is much reduced by the present RG calculation due to taking account of the non-logarithmic terms (the SB channels), which become important at high temperatures.
   It is expected that such a difference is further reduced by the modified-Kanamori approach\cite{NBTVT}, which is not applicable at low temperatures but gives better agreement with $\chi_{\rm QMC}(T)$ at high temperatures.
   In the present calculation, we take the linearized (continuum) dispersion, in which  lattice effects, coming from the use of the full tight binding spectrum $\epsilon_{p}(k) = -2t \left[\cos (k) - \cos (k_F ) \right]$, are neglected.
   When the lattice effects are considered in the calculation of $I_{\rm SB} (\ell )$, there appears a slight dip around $T\sim 0.1 t$\cite{NBTVT}.
   For $U=2V_2$ [plot (iii)] corresponding to the boundary between SDW and CDW, $G_\sigma$ is invariant due to $G_{\sigma}(0)=0$, so that $d\chi /d T$ is constant at low temperatures as for an ordinary paramagnetic metal.
   For $U < 2V_2$ [plot (iv)], $\chi$ is reduced to zero at $ T \ll t$  due to the CDW state.

   It is noticed that $\chi (T)$ is much enhanced by $V_1$ compared with $V_2$ at finite temperatures.
   %
   %
   This difference originates from the degree of the renormalization of $G_{4\perp}$.
   As seen in Fig. \ref{gflow}, in the presence of $V_1$ , the reduction of $G_{4\perp}$ [plot (ii')] due to the SB channel is smaller than that in the presence of $V_2$ [plot (iii')].
   Consequently, the enhancement of $\chi (T)$ by $V_1$ is much larger than by $V_2$ at finite temperatures.
   This superiority of $V_1$ is seen only at finite temperature.
   Such an effect of $V_1$ is surprising since $V_1$ was considered to be irrelevant whereas $V_2$ to be relevant\cite{Yoshioka}.

   The origin of this enhancement can be further analyzed by considering the  following random phase approximation (RPA) for $\chi(T)$  %
   \begin{align}
   \chi (T) =\frac{2}{\pi v_F}\frac{\chi_p^0 (T)}
   {1+\left[ G_\sigma (T) -G_{4\perp} (T) 
   \right] \chi_p^0 (T)/2},
   \label{RPA}
\end{align}
   where $\chi_p^0 (T)=\tanh (E_0 /4T) $ is the magnetic susceptibility of the non-interacting case per branch. 
   In Eq.(\ref{RPA}), $G_{\sigma (4\perp)}(T)$ denotes the renormalized coupling at $T$, i.e., $G_\sigma (T) = G_\sigma (0 ) /[1+G_\sigma (0)\ln (2T/E_0 )]$ and $G_{4\perp}(T)=[G_{4\rho}(\ell =0)-G_{4\sigma}(\ell =0)]/2$.
   Note that this formula includes the fluctuation beyond the simple RPA due to the renormalization of the Cooper and Peierls channel above $T$.
   The result of Eq. (\ref{RPA}) qualitatively agrees with that of Eq. (\ref{magsusRG}) except that it does not consider transients due to the interplay between the Cooper (or Peierls) and SB channel for $E_0 (\ell ) \sim 4T$.
   At $T=0$, Eq. (\ref{RPA}) yields $\chi (0) =2(\pi v_\sigma^*)^{-1}$ where $v_\sigma^*$ is given by
\begin{align}
   v_\sigma^* &\equiv 
    v_F \left[ 1 -G_{4\perp}/2\right] \nonumber\\
   &= v_F \left[
   1-(\pi v_F )^{-1}(U/2+V_1 +V_2 )
   \right].
   \label{spinvelo}
\end{align}
   As is clear from this form, $\chi (T)$ depends not only on $U$ but also on $V_1$ and $V_2$.
   In earlier studies, $\chi (0)$ is given merely by the inverse of $v_\sigma$, where the spin velocity $v_\sigma$ is expressed as $v_\sigma =v_F [1-U/(2\pi v_F )]$ and is independent of $V_1$ and $V_2$\cite{Solyom}.
   The present results reveal that $\chi (0)$ is not given by the ordinary spin-velocity $v_\sigma$, but by $v_\sigma^*$.
   Such a difference originates from a careful treatment of $H_{\rm I}$ [Eq. (\ref{HIrs})] (especially the $g_4$-process) and the accurate derivation of the flow of $z_\sigma$ [Eq. (\ref{dz})].
   It is worth remarking that in the rotationally invariant Hartree-Fock theory,  the $V_1$ interaction does contribute to the magnetic susceptibility.
   The corrections to the spin velocity obtained in this work are consistent with those of the Hartree-Fock theory. 
   We note that a recent result obtained by the numerical diagonalization also indicates the enhancement of $\chi(0)$ by $V_1$.\cite{TO}

	\begin{figure}[tb]
	\begin{center}\leavevmode
   \includegraphics[width=7cm]{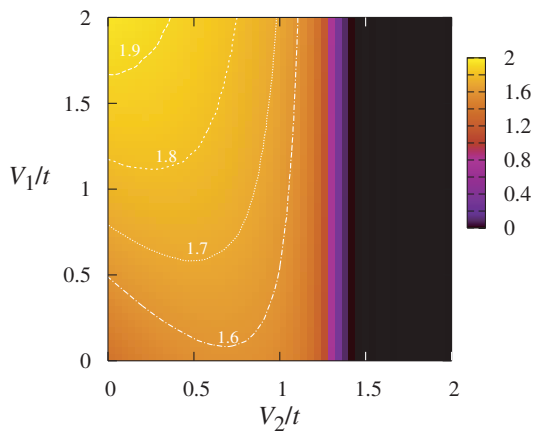}
	\end{center}
	\caption{
   Contour plot of $\chi(T)/\chi_0$ at $T=0.001t$ as a function of $V_1$ and $V_2$. The plot at $V_2 /t = U/2t=1$ corresponds to  the phase boundary between SDW and CDW.
   }
	\label{cont}
	\end{figure}
   %
   Figure \ref{cont} shows the contour plot of $\chi$ in  the $V_1$-$V_2$ plane at $T=0.001t$ and $U=2t$.
   It exhibits $\chi (T)$ is monotonically enhanced by increasing $V_1$.
   For large $V_2$, $\chi (T)$ decreases exponentially to zero with increasing $V_2$ indicating  the formation of CDW state.\cite{Solyom,Yoshioka}
   The $V_1$-$V_2$ dependence of $\chi (T)$ is approximately proportional to $V_1 + V_2 $ for small $V_2/t$.
   The magnitude of  $V_2/t (\simeq 1.3)$ for vanishing of $\chi$ is larger than  that of $T=0$, i.e.,  $V_2 = U/2 (=t)$, due to the effect of finite temperature. 

   The role of $V_1$ and  $V_2$ on $\chi$ can be intuitively understood as follows.
   First,  consider the Hubbard model ($U>0, V_1 =V_2 =0 )$, where the dominant state is the $2k_F$-SDW with a periodicity of four lattice spacing.
   The enhancement of $\chi (T)$ by $U (> 0)$ occurs due to the reduction of double occupancy of up spin and down spin.
   On the other hand,  at low temperatures  $\chi (T)$ is reduced due to irrelevance of the backward scattering.\cite{Solyom}
   Next, the addition of $V_1$, which prevents electrons from coming to the nearest-neighbor site, enhances the $2k_F$-SDW and the local moment of spin.
   This explains the enhancement of $\chi (T)$ in the presence of $V_1$ as seen from plots (1) and (2) in Fig. \ref{magsus}.
   When $V_2$ is added, electrons become less located at next-nearest-neighbor sites leading to the coexistence of $2k_F$-CDW and $2k_F$-SDW.
   The CDW fluctuations reduce  $\chi (T)$.

   Finally, the present results can be  discussed in connection with the experimental situation for the series of quasi-one-dimensional organic conductors (TMT$C$F)$_2 X$ ($C$=S, Se; $X$=PF$_6$, Br)\cite{Dumm}.
   In (TMTTF)$_2 X$ for example, $\chi (T)$ exhibits a maximum and $d\chi /dT$ is large at low temperatures.
   %
    The present results suggest that besides the effect of $U$, a moderate $V_1$ leads to a clear maximum in $\chi(T)$ and  to large $d\chi /dT$.
   Considering that a charge ordered state exists in (TMTTF)$_2 X$\cite{SHF}, these compounds should be characterized by a large $V_1$.
   Therefore the previous determinations of $U$ from $\chi (T)$ in (TMTTF)$_2 X$ might be considered as overestimations \cite{Dumm,Wzietek}.
   As for the Bechgaard salts, no charge ordered state is found in  (TMTSF)$_2$PF$_6$ and a much less pronounced maximum of $\chi (T)$ and smaller $d\chi /d T$ are observed.
   Besides the role of $U$, the present results would indicate  that (TMTSF)$_2$PF$_6$ has a relatively large $V_2$ (and not small $U$ and $V_1$).
   A large $V_2$ is consistent with a $2k_F$-SDW state coexisting with $2k_F$-CDW.\cite{PR,KOY} 
   %
   
   %
   %
   %
   %

   In conclusion, we calculated the magnetic susceptibility, $\chi (T)$, by considering the renormalization of the forward scattering in the same branch, the $g_4$-process, using the Kadanoff-Wilson RG technique.  
   With the careful treatment of the $g_4$-process, we found that $\chi (T)$ is enhanced by the nearest-neighbor interaction $V_1$, but is less enhanced by $V_2$. 
   It is also found that $\chi (0)$ is given by $v_\sigma^*$ [Eq. (\ref{spinvelo})], which is different from the ordinary spin-velocity $v_\sigma$.
   A comparison with the experimental situation of the quasi-one-dimensional organic compounds suggests that (TMTTF)$_2 X$ has a large $V_1$ ($\lesssim U$) but small $V_2$, while (TMTSF)$_2 X$ has a large $V_2$ ($\lesssim V_1 \lesssim U$).
   %

   The authors thank K. Kuroki, M. Ogata and Y. Tanaka for valuable discussions.
   The present work has been financially supported by a Grant-in-Aid for Scientific Research on Priority Areas of Molecular Conductors (No. 15073103 and 15073213) from the Ministry of Education, Culture, Sports, Science and Technology, Japan.

\end{document}